\documentclass[pra,twocolumn]{revtex4}
\usepackage{amssymb}
\usepackage{amsfonts}
\usepackage{amsmath}
\usepackage{graphicx}
\usepackage{bm}
\usepackage{braket}

\begin{document}

\title{Energy-level inversion for vortex states in spin-orbit coupled
Bose-Einstein condensates}
\author{Huan-Bo Luo$^{1,2}$}
\author{Lu Li$^3$}
\author{Boris A. Malomed$^{4,5}$}
\author{Yongyao Li$^{1,2}$}
\author{Bin Liu$^{1,2}$ }
\email{binliu@fosu.edu.cn}
\affiliation{$^1$School of Physics and Optoelectronic Engineering, Foshan University,
Foshan 528000, China}
\affiliation{$^2$Department of Physics, South China University of Technology, Guangzhou 510640, China}
\affiliation{$^3$Institute of Theoretical Physics and Department of Physics, Shanxi
University, Taiyuan 030006, China}
\affiliation{$^4$Department of Physical Electronics, School of Electrical Engineering,
Faculty of Engineering, Tel Aviv University, Tel Aviv 69978, Israel}
\affiliation{$^5$Instituto de Alta Investigaci\'{o}n, Universidad de Tarapac\'{a},
Casilla 7D, Arica, Chile}

\begin{abstract}
We investigate vortex states in Bose-Einstein condensates under the combined
action of the spin-orbit coupling (SOC), gradient magnetic field, and
harmonic-oscillator trapping potential. The linear version of the system is
solved exactly. Through the linear-spectrum analysis, we find that, varying
the SOC strength and magnetic-field gradient, one can perform energy-level
inversion. With suitable parameters, initial higher-order vortex states can
be made the ground state (GS). The nonlinear system is solved numerically,
revealing that the results are consistent with the linear predictions in the
case of repulsive inter-component interactions. On the other hand,
inter-component attraction creates the GS in the form of mixed-mode states
in a vicinity of the GS phase-transition points. The spin texture of both
vortex- and mixed-mode GSs reveals that they feature the structure of 2D (%
\textit{baby}) skyrmions.
\end{abstract}

\maketitle


\section{Introduction}

Atomic Bose-Einstein condensates~(BECs) are versatile platforms for
simulations of various phenomena from condensed-matter physics \cite%
{RepProgPhys.75.082401,Lewenstein}. Among these phenomena, the spin-orbit
coupling (SOC) plays a basic role in spin Hall effects~\cite%
{RevModPhys.82.1959}, topological insulators~\cite{RevModPhys.82.3045},
spintronic devices~\cite{RevModPhys.76.323}, etc. The experimental
realization of SOC in one-dimensional~(1D) \cite{nature09887,Juzeliunas} and
two-dimensional~(2D)~\cite{Science.354.83} two-component BEC has inspired
theoretical research into spin-orbit-coupled BECs. The analysis based on the
Gross-Pitaevskii equations (GPEs) has produced remarkable phenomena, such as
vortices \cite{Kawakami,Drummond,
Sakaguchi,PhysRevE.89.032920,PhysRevE.94.032202,CNSNS}, solitons \cite%
{PhysRevLett.110.264101,1D sol 2,1D sol 3,1D sol 4, Cardoso,Lobanov,2D SOC
gap sol Raymond,SOC 2D gap sol Hidetsugu,low-dim SOC,Han Pu 3D}, and
skyrmions \cite{PhysRevLett.109.015301}. Comprehensive insights into
experimental and theoretical achievements in this field are provided by
review \cite{Spielman,Galitski,Ohberg,Zhai,SOC-sol-review,Sherman}.

2D solitons supported by the interplay of SOC and cubic attractive
interactions in the free space are remarkable modes due to their stability
against the collapse and specific vortex structure: depending on the
relative strength of the cross- and self-attraction, stable modes are
\textit{semi-vortices}, with vorticities $0$ and $1$ in the two components,
and \textit{mixed modes}, which include terms with vorticities $\left(
0,+1\right) $ in one component, and $\left( -1,0\right) $ in the other \cite%
{PhysRevE.89.032920,PhysRevE.94.032202}. However, semi-vortex and mixed-mode
solitons are stable only when they represent the ground state (GS), while
the corresponding excited states, produced by the addition of the equal
vorticities to both components, are completely unstable. Stabilization of
the higher-vorticity states can be provided if a tunable mechanism of
energy-level inversion can be introduced, that modifies the eigen-energy
spectrum while preserving the corresponding eigenfunctions. Ideally, it
should enable the transformation of any higher-order vortex state into the
GS, thereby paving the way for the experimental realization. Recently, a
possibility of the transformation of any excited state into the respective
GS in the 1D BEC with SOC and gradient magnetic field has been demonstrated
in Ref.~\cite{PhysRevA.106.063311}.

In this paper, we introduce the 2D SOC system which includes the gradient
magnetic field and the harmonic-oscillator (HO) trapping potential. The
linear version of the system is solved exactly. The solution demonstrates
that the combined effect of SOC and magnetic field leads to reduction of the
total energy of the vortex states, the size of the effect growing with the
increase of the vorticity. Thus, by adjusting the SOC strength and
magnetic-field gradient, one can realize the energy level inversion, making
it possible to convert any higher-order vortex state into the GS. The full
nonlinear system, including either repulsive or attractive inter-component
interaction, is solved numerically.

The following presentation is structured as follows. The model is introduced
in Sec. II. The linear solution is constructed in Sec. III, in terms of wave
functions of Landau levels. In Secs. IV , numerical solutions of the
nonlinear system with inter-component repulsion or attraction are produced.
In Sec. V, spin textures of the newly found two-component states are
presented in detail. The paper is concluded by Sec. VI.

\section{The model}

We consider the spin-orbit-coupled effectively 2D binary BEC under the
action of the HO trapping potential, written in the scaled form as $%
V=r^{2}/2 $, and dc magnetic field $\mathbf{B}=(-\alpha x,-\alpha y,\Omega )$%
, which has a constant gradient $-\alpha $ along the $x$ and $y$
directions,while its $z$-component $\Omega $ is constant. The Rashba SOC is
represented by operator $V_{\text{so}}=i\beta (\sigma _{y}\partial
_{x}-\sigma _{x}\partial _{y})$ in the system of GPEs for the binary BEC,
where $\mathbf{\sigma }=(\sigma _{x},\sigma _{y},\sigma _{z})$ is the vector
of the Pauli matrices, and $\beta $ is the SOC\ strength. The scaled GPE
system for the spinor wave function, $\Psi =(\Psi _{1},\Psi _{2})^{T}$, is
\begin{equation}
\begin{split}
i\partial _{t}\Psi _{1}=& \frac{1}{2}\left( -\nabla ^{2}+r^{2}\right) \Psi
_{1}+\Omega \Psi _{1}-\alpha \left( x-iy\right) \Psi _{2} \\
& +\beta \left( \partial _{x}-i\partial _{y}\right) \Psi _{2}+\left( g|\Psi
_{1}|^{2}+\gamma |\Psi _{2}|^{2}\right) \Psi _{1}, \\
i\partial _{t}\Psi _{2}=& \frac{1}{2}\left( -\nabla ^{2}+r^{2}\right) \Psi
_{2}-\Omega \Psi _{2}-\alpha \left( x+iy\right) \Psi _{1} \\
& -\beta \left( \partial _{x}+i\partial _{y}\right) \Psi _{1}+\left( g|\Psi
_{2}|^{2}+\gamma |\Psi _{1}|^{2}\right) \Psi _{2},
\end{split}
\label{main}
\end{equation}%
where $\Omega $ plays the role of the effective Zeeman splitting between the
components, $g$ and $\gamma $ are coefficients of the intra- and
inter-component interactions, respectively~\cite{2D-magnetic-1,2D-magnetic-2,
2D-magnetic-3}. Using the remaining scaling invariance
of Eq.~\eqref{main}, we set $g=1$, assuming the repulsive sign of the
self-interaction in each component.\ However, $\gamma $ may be negative,
representing the possibility of the attraction between the components, which
can be introduced by means of the Feshbach resonance~\cite{Feshbach}. For
the use in the following analysis, we denote%
\begin{equation}
\Omega \equiv \beta\Delta-\frac{1}{2}  \label{Delta},
\end{equation}%
where $\Delta$ serves as a tunable parameter employed to control the strength 
of Zeeman splitting.

Stationary solutions of Eq.~\eqref{main} with chemical potential $\mu $ are
sought for in the usual form,
\begin{equation}
\Psi (x,y,t)=\exp (-i\mu t)\{\psi _{1}(x,y),\psi _{2}(x,y)\}^{T},
\label{psipsi}
\end{equation}%
with functions $\psi _{1,2}(x,y)$ satisfying equations
\begin{equation}
\begin{split}
\mu \psi _{1}=&\frac{1}{2}\left( -\nabla ^{2}+r^{2}\right) \psi _{1}+
\Omega \psi _{1}-\alpha \left( x-iy\right) \psi_{2} \\
&+\beta \left( \partial _{x}-i\partial _{y}\right) \psi _{2}+\left(g|\psi
_{1}|^{2}+\gamma |\psi _{2}|^{2}\right) \psi _{1}, \\
\mu \psi _{2}=&\frac{1}{2}\left( -\nabla ^{2}+r^{2}\right) \psi _{2}-
\Omega \psi _{2}-\alpha \left( x+iy\right) \psi_{1} \\
&-\beta \left( \partial _{x}+i\partial _{y}\right) \psi _{1}+\left( g|\psi
_{2}|^{2}+\gamma |\psi _{1}|^{2}\right) \psi _{2}~.
\end{split}
\label{psi12}
\end{equation}

The SOC system (\ref{main}) conserves \ two evident dynamical invariants,
\textit{viz}., the total norm of the two components,%
\begin{equation}
N=\int \int \left[ \left\vert \psi _{1}\left( x,y\right) \right\vert
^{2}+\left\vert \psi _{2}\left( x,y\right) \right\vert ^{2}\right] dxdy,
\label{N}
\end{equation}%
and energy (Hamiltonian),
\begin{equation}
\begin{split}
    E=&\iint  \left\{ \frac{1}{2}\!\left(\left\vert \nabla \psi _{1}\right\vert
    ^{2}\!+\!\left\vert \nabla \psi _{2}\right\vert ^{2}\right)\!+\!\frac{r^{2}}{2}\!\left( \left\vert \psi
    _{1}\right\vert ^{2}+\left\vert \psi _{2}\right\vert ^{2}\right)
    \right.   \\
    &-\alpha \left[ \left( x-iy\right) \psi _{1}^{\ast }\psi _{2}+\left( x+iy\right)\psi
    _{1}\psi _{2}^{\ast } \right]  \\
    &-\beta \left[ \psi _{2}\left( \partial _{x}-i\partial _{y}\right) \psi
    _{1}^{\ast }+\psi _{2}^{\ast }\left( \partial _{x}+i\partial _{y}\right)
    \psi _{1}\right]  \\
    &+\Omega \left( \left\vert \psi
    _{1}\right\vert ^{2}-\left\vert \psi _{2}\right\vert ^{2}\right)+\gamma \left\vert \psi _{1}\right\vert
    ^{2}\left\vert \psi _{2}\right\vert ^{2}    \\
    &\left. +\frac{g}{2}\left( \left\vert \psi _{1}\right\vert ^{4}+\left\vert
        \psi _{2}\right\vert ^{4}\right)\right\} dxdy,
\end{split} \label{E}
\end{equation}
where $\ast $ stands for the complex conjugate. Note that the terms $\sim
\beta $ in expression (\ref{E}), which seem formally asymmetric with respect
to $\Psi _{1}$ and $\Psi _{2}$, are actually symmetric, if one takes into
regard the possibility of the application of integration by parts, $\Psi
_{2}\left( \partial _{x}-i\partial _{y}\right) \Psi _{1}^{\ast }\rightarrow
-\Psi _{1}^{\ast }\left( \partial _{x}-i\partial _{y}\right) \Psi _{2}$. The
system's GS corresponds to a minimum of the energy for a fixed value of the
norm. In fact, for the linearized system, with $g=\gamma =0$, the energy of
stationary states is proportional to the respective chemical potential,
determined by the stationary GPE system (\ref{psi12}), therefore, in the
linear limit, the energy minimization is tantamount to the minimization of $%
\mu $.

We note that, in terms of polar coordinates $\left( r,\theta \right) $ and
wave-function components
\begin{equation}
\tilde{\Psi}_{1}\equiv \Psi _{1},\tilde{\Psi}_{2}\equiv e^{-i\theta }\Psi
_{2},  \label{tilde}
\end{equation}%
expression (\ref{E}) for the energy takes an \emph{axisymmetric} form:%
\begin{equation}
\begin{split}
E=&\int_{0}^{\infty }rdr\int_{0}^{2\pi }d\theta \left\{ %
\frac{1}{2}\sum_{j=1,2}\left( \left\vert \frac{\partial \tilde{\psi}_{j}}{\partial r}%
\right\vert ^{2}+\frac{1}{r^{2}}\left\vert \frac{\partial \tilde{\psi}_{j}}{%
\partial \theta }\right\vert ^{2}\right) \right.   \\
&+\frac{r^{2}}{2}\left(\left\vert \tilde{\psi}%
_{1}\right\vert ^{2}+\left\vert \tilde{\psi}%
_{2}\right\vert ^{2}\right)-\alpha r\left( \tilde{\psi}_{1}^{\ast }\tilde{\psi}_{2}+\tilde{\psi}_{1}%
\tilde{\psi}_{2}^{\ast }\right)   \\
&-\beta \left[ \tilde{\psi}_{2}\left( \frac{\partial }{\partial r}-\frac{i}{r}%
\frac{\partial }{\partial \theta }\right) \tilde{\psi}_{1}^{\ast }+\tilde{%
\psi}_{2}^{\ast }\left( \frac{\partial }{\partial r}+\frac{i}{r}\frac{%
\partial }{\partial \theta }\right) \tilde{\psi}_{1}\right]   \\
&+ \Omega\left(\left\vert \tilde{\psi}_{1}\right\vert ^{2}-\left\vert
\tilde{\psi}_{2}\right\vert ^{2}\right) +\gamma \left\vert \tilde{\psi}_{1}\right\vert
^{2}\left\vert \tilde{\psi}_{2}\right\vert ^{2} \\
&\left.+\frac{g}{2}\left(\left\vert \tilde{\psi%
}_{1}\right\vert ^{4}+\left\vert \tilde{\psi%
}_{2}\right\vert ^{4}\right)\right\} .  \label{E-tilde}
\end{split}%
\end{equation}
The invariance of energy (\ref{E-tilde}) with respect to the rotation by
arbitrary angle, $\theta \rightarrow \theta +\delta \theta $, implies that
the additional dynamical invariant of the system is the angular momentum,
which is defined as
\begin{equation}
    \begin{split}
M=&\int \int \left( \tilde{\psi}_{1}^{\ast }\hat{L}\tilde{\psi}_{1}+\tilde{%
\psi}_{2}^{\ast }\hat{L}\tilde{\psi}_{2}\right) dxdy  \notag \\
\equiv& \int \int \left[ \psi _{1}^{\ast }\hat{L}\psi _{1}+\psi _{2}^{\ast
}\left( \hat{L}-1\right) \psi _{2}\right] dxdy,
\end{split}\label{momentum}
\end{equation}
where $\hat{L}=i\left( y\partial _{x}-x\partial _{y}\right) \equiv
-i\partial _{\theta }$ is the canonical angular-momentum operator.

It is relevant to mention that essentially the same GPE system (\ref{main})
can be derived if, instead of the real magnetic field, a synthetic field is
used. It is well known that synthetic magnetic fields can be induced\ by
rapid rotation of the condensate \cite{Cornell}, or by an appropriate
combination of illuminating laser beams \cite{Spielman-2}. The use of
synthetic fields opens the way to realization of many fascinating phenomena,
such as the Dirac's monopole \cite{monopole}.

Equations~\eqref{main} and (\ref{psi12}) are written in the scaled form. In
physical units, assuming that the binary condensate is a mixture of two
different atomic states of $^{87}$Rb \cite{nature09887}, a relevant value of
the HO trapping frequency is $\omega =10$ Hz. The number of atoms in the
condensates is $1000$, which is sufficient for the experimental observation
of the predicted patterns in full detail. The characteristic length, time
and energy are identified as $l=\sqrt{\hbar /m_{\mathrm{at}}\omega }=8.55$ $%
\mathrm{\mu }$m, $\tau =1/\omega =100$ ms, and $\epsilon =\hbar \omega
=1.05\times 10^{-33}$ J, where $m_{\mathrm{at}}=1.44\times 10^{-25}$ kg is
the atomic mass of $^{87}$Rb. The strength of SOC, denoted by 
$\beta=l\pi/\left(\sqrt{3}\lambda\right)$, where $\lambda$ represents the 
wavelength of the laser, can be adjusted across a wide range depending on the specific 
configurations of the laser system \cite{BEC-SOC GP eqns}. Moreover, the shorter the wavelength of the 
laser, the greater the SOC strength. For instance, the Nd:YAG lasers typically 
emit light with a wavelength of 1064 nm, corresponding to a SOC strength of 
$\beta=1.47$, while the He-Ne lasers emit light with a wavelength of 633 nm, 
resulting in a higher SOC strength of $\beta=2.45$. 

\section{The solution of the linear system}

We first solve the linear version of Eq.~\eqref{psi12}, i.e.,
\begin{equation}
\hat{H}\psi \!=\!\mu \psi ,  \label{eigen}
\end{equation}%
where the Hamiltonian can be represented in the compact form,
\begin{equation}
\hat{H}\!=\!%
\begin{bmatrix}
\hat{a}^{\dagger }\hat{a}+\hat{b}^{\dagger }\hat{b}+\frac{1}{2}+\beta \Delta
& (\beta -\alpha )\hat{b}-(\alpha +\beta )\hat{a}^{\dagger } \\
(\beta -\alpha )\hat{b}^{\dagger }-(\alpha +\beta )\hat{a} & \hat{a}^{\dag }%
\hat{a}+\hat{b}^{\dag }\hat{b}+\frac{3}{2}-\beta \Delta%
\end{bmatrix}%
,
\end{equation}%
where the creation and annihilation operators are introduced as $\hat{a}%
^{\dagger }=(x-iy-\partial _{x}+i\partial _{y})/2$, $\hat{a}=(x+iy+\partial
_{x}+i\partial _{y})/2$, $\hat{b}^{\dagger }=(x+iy-\partial _{x}-i\partial
_{y})/2$ and $\hat{b}=(x-iy+\partial _{x}-i\partial _{y})/2$. To solve the
linear stationary Schr\"{o}dinger equation (\ref{eigen}), a series of wave
functions of the Landau levels are introduced as the basis \cite{LL}:
\begin{equation}
\begin{split}
& f_{n,m}(r,\theta ) \\
& =\frac{\exp \left( im\theta -r^{2}/2\right) }{\sqrt{\pi n!(n+m)!}}%
\sum_{k=0}^{n+m}C_{n+m}^{k}A_{n}^{k}(-1)^{k}r^{2(n-k)+m},
\end{split}
\label{Landau}
\end{equation}%
where $m$ is the winding number (alias vorticity, or magnetic quantum
number), $n$ is an auxiliary quantum number, and $n+m$ is the Landau-level
index. The ranges of $n$ and $m$ are $n=0,1,2,\ldots $ and $%
m=-n,-n+1,-n+2,\ldots $, respectively ($m$ also takes positive integer
values). $C_{n}^{m}=\frac{n!}{m!(n-m)!}$ are
the binomial coefficients, and $A_{n}^{m}=\frac{n!}{(n-m)!}$ for $m\leq n$; $%
A_{n}^{m}\equiv 0$ for $m>n$. The action of operators $\hat{b}^{\dagger }$
and $\hat{a}^{\dagger }$ on wave function (\ref{Landau}) amounts to
\begin{equation}
\begin{array}{ll}
\hat{a}^{\dagger }f_{n,m}=\sqrt{n+1}f_{n+1,m-1},& \hat{a}f_{n,m}=\sqrt{n}%
f_{n-1,m+1}, \\
\hat{b}^{\dagger }f_{n,m}=\sqrt{n+m+1}f_{n,m+1},& \hat{b}f_{n,m}=\sqrt{n+m}%
f_{n,m-1}.
\end{array}
\label{rise_down}
\end{equation}

In the case of equal SOC and magnetic-field-gradient coefficients in Eqs. (%
\ref{main}) and (\ref{psi12}), $\alpha =\beta $, the present linear system~(\ref{eigen})
admits an exact solution (eigenstate), in terms of wave functions (\ref%
{Landau}),
\begin{equation}
\begin{split}
\psi _{1}^{n,m}& =\frac{1}{\sqrt{\mathcal{N}_{n}}}f_{n,m}(r,\theta ), \\
\psi _{2}^{n,m}& =\frac{\sqrt{4n+\Delta ^{2}}+\Delta }{2\sqrt{\mathcal{N}%
_{n}n}}f_{n-1,m+1}(r,\theta ),
\end{split}
\label{eigenfunction}
\end{equation}%
where\ $\Delta $ is defined by Eq.~(\ref{Delta}), and the normalization
coefficient is
\begin{equation}
\mathcal{N}_{n}=%
\left\{\begin{array}{ll}
1, & n=0, \\
\displaystyle\frac{4n+\Delta ^{2}+\Delta \sqrt{4n+\Delta ^{2}}}{2n}, &
n=1,2,3,\ldots ~.%
\end{array}\right.
\end{equation}%
The difference $\delta m=1$ between the two components of eigenstate (\ref%
{eigenfunction}) is a characteristic feature of bound states of the
above-mentioned semi-vortex type supported by SOC in the 2D space \cite%
{PhysRevE.89.032920}. The corresponding eigenvalues of the chemical
potential are
\begin{equation}
\mu _{n,m}=\left\{ %
\begin{array}{ll}
\frac{1}{2}+m+\beta \Delta , & n=0, \\
\frac{1}{2}+2n+m-\beta \sqrt{4n+\Delta ^{2}}, & n=1,2,3,\ldots ~.%
\end{array}%
\right.  \label{eigenvalue}
\end{equation}%
For given quantum number $n$, the eigenvalues (\ref{eigenvalue}) attain a
minimum at $m=-n$, namely,%
\begin{equation}
\mu _{n,m=-n}\equiv \mu _{n}=\left\{
\begin{array}{ll}
\frac{1}{2}+\beta \Delta , & n=0, \\
\frac{1}{2}+n-\beta \sqrt{4n+\Delta ^{2}}, & n=1,2,3,\ldots ~%
\end{array}%
\right.  \label{mu_n}
\end{equation}%
(however, this conditional minimum, corresponding to a particular value of $%
n $, does not imply the system's GS, which should be identified as the
absolute minimum). The respective\ wave function is%
\begin{equation}
    \begin{split}
    \psi _{1}^{n,m=-n}& =R_{1}^{(n)}(r)\exp(-in\theta ), \\
    \psi _{2}^{n,m=-n}& =R_{2}^{(n)}(r)\exp\left[ -i(n-1)\theta \right],
    \end{split}
    \label{m=-n}
\end{equation}%
where the radial wave functions $R_{1,2}^{(n)}$ for $n\geq 1$ are
\begin{equation}
\begin{split}
R_{1}^{(n)}& =\frac{1}{\sqrt{\pi n! \mathcal{N}_{n}}}r^{n}\exp \left( -\frac{%
r^{2}}{2}\right) , \\
R_{2}^{(n)}& =\frac{\sqrt{4n+\Delta ^{2}}+\Delta }{2\sqrt{\pi n!\mathcal{N}%
_{n}}}r^{{n}-1}\exp \left( -\frac{x^{2}}{2}\right) ,
\end{split}
\label{R}
\end{equation}%
and for $n=0$,
\begin{equation}
R_{1}^{(0)}=(1/\sqrt{\pi })\exp (-r^{2}/2),~R_{2}^{(0)}=0.  \label{0}
\end{equation}%
Typical profiles of the radial wave functions are plotted in Figs.~\ref%
{figure1}(a-c).

\begin{figure}[tbp]
\centering
\includegraphics[width=3.4in]{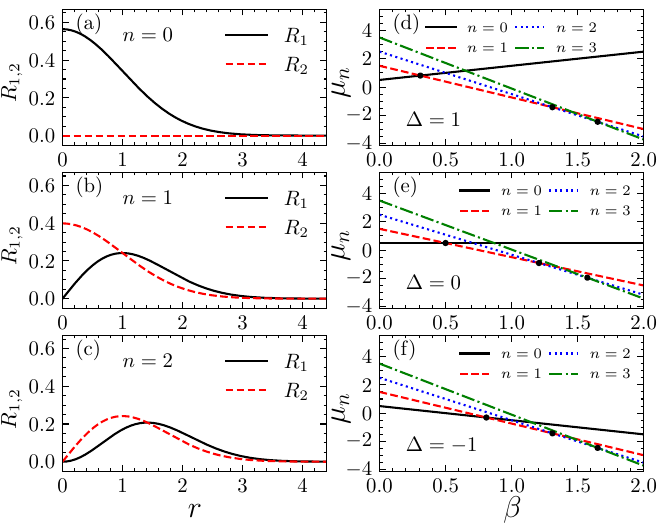}
\caption{(a)-(c): Profiles of the radial wave functions $R_{1,2}^{(n)}(r)$,
defined as per Eqs. (\protect\ref{R}) and (\protect\ref{0}), with quantum
numbers (a) $n=0$, (b) $n=1$ and (c) $n=2$. (d)-(f): The corresponding
chemical potential $\protect\mu _{n}(\protect\beta )$ with (d) $\Delta =1$,
(e) $\Delta =0$, and (f) $\Delta =-1$, plotted pursuant to Eq.~\eqref{mu_n}.
The dots are values of $\protect\beta _{n}$ defined by Eq.~\eqref{beta_n}.}
\label{figure1}
\end{figure}

Thus, the quantum number $n=-m$ can be used to label the order of vortex
states (\ref{m=-n}). It is seen that the SOC strength $\beta $ alters the
spectrum~\eqref{mu_n} of eigenvalues $\mu _{n}$, but does not affect the
corresponding eigenfunctions~\eqref{R}. In particular, Figs.~\ref{figure1}%
(d-f) display the dependence of $\mu _{n}$ (alias energy, as it is
proportional to the chemical potential for the linearized system) on $\beta $
at $\Delta =0$ and $\pm 1$. Branches $\mu _{n}$ with larger numbers $n$ vary
faster as functions of $\beta $. The intersection of the branches
corresponding to values $n$ and $n+1$ implies the GS switching. The
respective critical values $\beta _{n}$ at the switching points can be
obtained by solving equation $\mu _{n}(\beta _{n})=\mu _{n+1}(\beta _{n})$:

\begin{equation}
\beta _{n}\!=\!\frac{1}{4}\left\{
\begin{array}{ll}
\vspace{0.1cm}\displaystyle\sqrt{4+\Delta ^{2}}-\Delta , & n=0, \\
\displaystyle\sqrt{4n+4+\Delta ^{2}}+\sqrt{4n+\Delta ^{2}}, & n=1,2,\ldots ~.%
\end{array}%
\right.  \label{beta_n}
\end{equation}%
Thus we find the fact that vortex states with any order $n$ can become the
system's GS by adjusting the value of the SOC strength $\beta $.

The above analysis is based on the special case of $\alpha =\beta $ (equal
SOC and magnetic-field-gradient strengths). The analysis can be extended for
$\alpha \neq \beta $, setting $\Delta =0$ in Eq. (\ref{eigen}), i.e., $%
\Omega =-1/2$ in Eq. (\ref{main}). Then, an approximate solution of Eq.~%
\eqref{eigen} with winding number $m$ (alias magnetic quantum number) can be
looked for as a combination of the Landau-level wave functions truncated at $%
n=N_t$:
\begin{equation}
\begin{split}
\psi _{1}& =\sum_{n=0}^{N_t}c_{n}f_{n,m}, \\
\psi _{2}& =\sum_{n=0}^{N_t}d_{n}f_{n-1,m+1},
\end{split}
\label{general}
\end{equation}%
where $c_{n}$ and $d_{n}$ are coefficients to be determined. We here produce
the approximate results for $N_t=50$, which are practically exact. By
substituting the ansatz~\eqref{general} into Eq.~\eqref{eigen} , we derive a
set of coupled linear equations for $c_{n}$ and $d_{n}$:
\begin{equation}
\begin{split}
\mu c_{n}& =\left( 2n+m+\frac{1}{2}\right) c_{n}-(\alpha +\beta )\sqrt{n}%
d_{n} \\
& +(\beta -\alpha )\sqrt{n+m+1}d_{n+1}, \\
\mu d_{n}& =\left( 2n+m+\frac{1}{2}\right) d_{n}-(\alpha +\beta )\sqrt{n}%
c_{n} \\
& +(\beta -\alpha )\sqrt{n+m}c_{n-1}.
\end{split}
\label{couple}
\end{equation}%
Once the quantum number $m$ is fixed, Eq.~\eqref{couple} can be solved by
dint of numerical diagonalization of the corresponding matrix. By comparing
the chemical potentials $\mu $ corresponding to different winding numbers $m$%
, we can thus identify the system's GS.

\begin{figure}[tbp]
\centering
\includegraphics[width=3.4in]{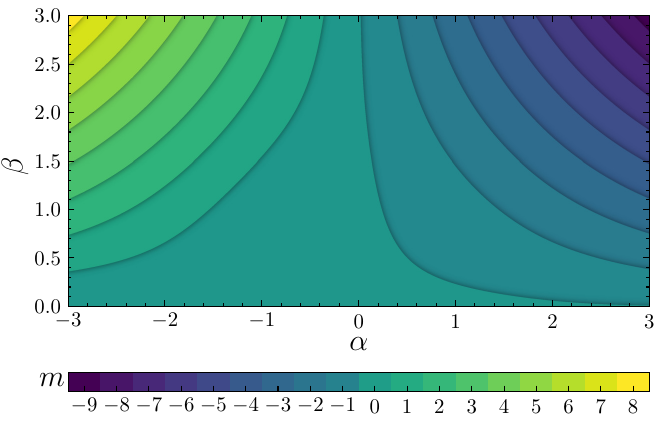}
\caption{The map of values of the winding number (magnetic quantum number) $%
m $ corresponding to GS of the linear system in the ($\protect\alpha ,%
\protect\beta $) parameter plane.}
\label{figure2}
\end{figure}

The so produced map of quantum numbers $m$, corresponding to the GS, in the $%
(\alpha ,\beta )$ parameter plane (alias the GS phase diagram) is plotted in
Fig.~\ref{figure2}. Along the diagonal, $\alpha =\beta $, the GS predicted
by this diagram agrees with the exact one given by Eq.~\eqref{beta_n} (note
the symmetry about the diagonal). The phase diagram demonstrates that the GS
switching between quantum numbers $m$ and $m+1$ takes place at $\alpha \neq
\beta $ as well. Furthermore, we find that if fixing $\beta=3.0$, as 
$\alpha$ changes from negative to positive, the quantum number $m$ gradually 
transitions from positive to negative. Due to the symmetric structure of states 
with quantum numbers $m$ and $-m$, we focused only on the case where $m \leq 0$,
or in other words, the case where $\alpha \geq 0$.

\section{The numerical solution for the nonlinear system}

We address the complete form of Eq.~\eqref{psi12}, including the nonlinear
terms, repulsive or attractive, while fixing $\alpha =\beta $, as in the
exact solution of the linear system. While energy-level inversion can be
achieved by changing both the parameters $\Delta$ and $\beta$, altering
$\Delta$ also changes the wave functions of each energy level,
whereas altering $\beta$ does not affect the wave functions. To focus on
the manipulations of the spectrum,
in the following discussion, we vary $\beta$ while keeping $\Delta = 0$. In
this case, stationary states can be found in the numerical form by means of
the imaginary-time propagation method \cite{Bao}, fixing the total norm of
the solution as $N=1$, see Eq. (\ref{N}). The input which was used to generate
the solutions by means
of this method was taken as a superposition of the vortex
components of the linear eigenmodes given by Eqs. (17)
and (18), \textit{viz}.,
\begin{equation}
    \psi_1=\psi_2=\sum_{n=0}^{N_s}\frac{r^n\exp(-in\theta-r^2/2)}{\sqrt{2\pi N_s}}.
    \label{input}
\end{equation}
For numerical calculations with $0\leq\beta\leq2$, choosing $N_s=5$ in Eq.~\eqref{input} is sufficiently
to generate stable eigenmodes of the nonlinear system.

We dwell on four cases, $\gamma
=1 $, $\gamma =0$, $\gamma =-1$, and $\gamma =-2$, which correspond, respectively, to the
repulsive, zero, attraction and stronger attraction between the
components. Note that the commonly known miscibility condition for the
binary Bose gas in the free space is, in the present notation, $\gamma <1$
\cite{Mineev}. The case of $\gamma =1$ corresponds to the miscibility
boundary, but the pressure if the OH trapping potential induces effective
miscibility in this case.

\begin{figure}[tbp]
\centering
\includegraphics[width=3.4in]{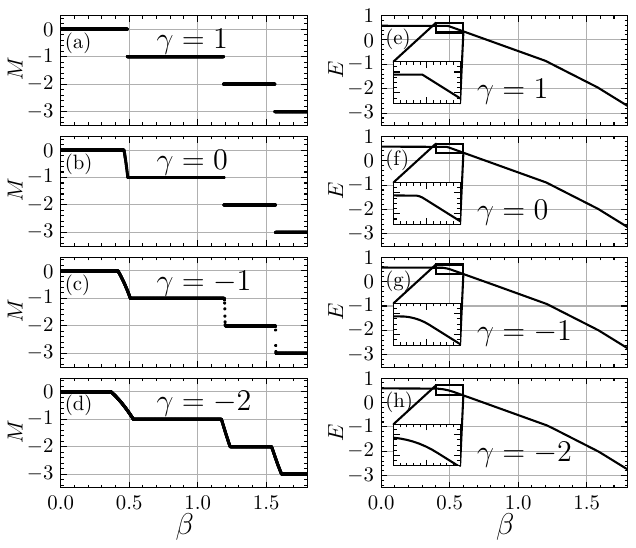}
\caption{(a-d) Angular momentum $M$ and (e-h) energy $E$, defined
as per Eqs.~\eqref{momentum} and~\eqref{E}, respectively, as
produced by the imaginary-time simulations of the full (nonlinear) system (%
\protect\ref{main}), for $\protect\beta $ varying from $0$ to $1.8$, with
step $\delta \protect\beta =0.001$, at (a, e) $\protect\gamma =1$, (b, f) $\protect%
\gamma =0$, (c, g) $\protect\gamma =-1$ and (d, h) $\protect\gamma =-2$.}
\label{figure3}
\end{figure}

The stationary nonlinear states may be naturally quantified by the angular
momentum (\ref{momentum}). It is easy to check $M=m$ if one substitutes
eigenfunction~\eqref{eigenfunction} into Eq.~\eqref{momentum}. Thus, integer
values of $M$ indicate vortex (or semi-vortex \cite{PhysRevE.89.032920})
states, while non-integer values of $M$ indicate mixed-mode states, defined
as in Ref. \cite{PhysRevE.89.032920}. The dependence of $M$ on $\beta $,
produced by the numerical solution is shown in Fig.~\ref{figure3}.

\begin{figure}[tbp]
\centering
\includegraphics[width=3.4in]{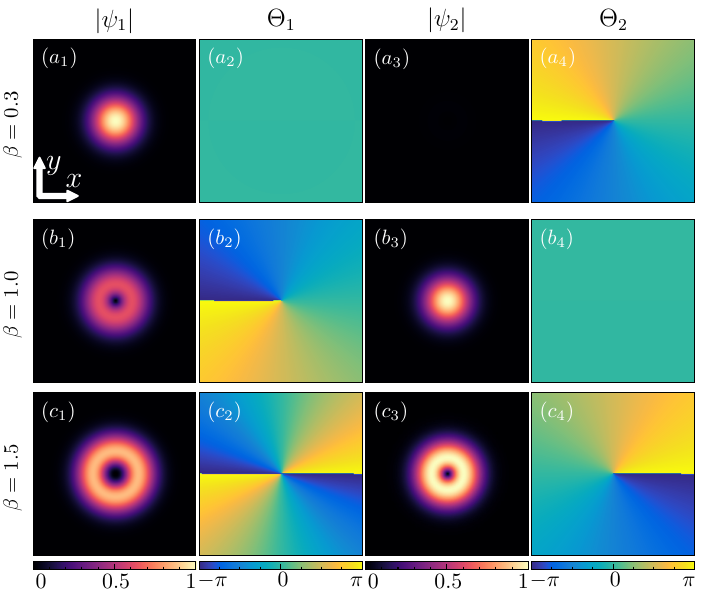}
\caption{Distributions of the absolute values, $\left\vert \protect\psi %
_{1,2}\right\vert $, and phases, $\Theta _{1,2}$, of wave functions of the
two components in the GS of the (semi-) vortex type, for $\protect\beta =0.3$
(panels $a_{1}-a_{4}$, with vorticities $0$ and $+1$ in the two componets;
panel $a_{3}$ seems as an empty one, as $\left\vert \protect\psi %
_{2}\right\vert $ is very small in that case); $\protect\beta =1.0$ (panels $%
b_{1}-b_{4}$, with vorticities $-1$ and $0$); and $\protect\beta =1.5$
(panels $c_{1}-c_{4}$, with vorticities $-2$ and $-1$). Other parameters in
Eq.~\eqref{main} are $\Delta =0$ and $\protect\gamma =1$.}
\label{figure4}
\end{figure}

For the different values of $\gamma $, dependences $M(\beta )$ in Fig.~\ref%
{figure3} (a-d) exhibit similar patterns, except in the vicinity of the
phase-transition points. Recall that the GS phase transitions in the exact
solution for the linear system are given by Eq.~\eqref{beta_n} -- in
particular, $\beta _{0}=0.5$, $\beta _{1}=1.2$, and $\beta _{2}=1.6$. As $%
\beta $ increases, $M$ follows the descending staircase-like pattern,
dropping by $1$ while passing each phase transition. Flat segments of the $%
M(\beta )$ dependences are populated by the (semi-)vortex two-component
states, see typical examples in Fig.~\ref{figure4}, while the oblique
transition segments carry mixed-mode states, see examples in Fig.~\ref%
{figure5}. Parameter $\gamma $ affects the width of the phase-transition
region, which is wider for smaller value of $\gamma $, in agreement with the
general trend to stabilization of mixed modes and destabilization of
semi-vortices following the decrease of $\gamma $ (increase of $-\gamma $)
\cite{PhysRevE.89.032920}.

Curves for energy $E$ as a function of $\beta$ are plotted in
Fig.~\ref{figure3} (e-h). We can find that as $\gamma$ decreases, 
the curves near the phase transition point gradually become smoother.
For regions far from the phase transition point, these curves are similar 
for different values of $\gamma$. Energy $E$ gradually decreases with the 
increase of $\beta$. Furthermore, each time that $\beta$ passes a phase-transition
point, the rate of energy reduction accelerates. These
results are similar to those for the dependences $\mu_n(\beta)$
of the chemical potenital displayed in Eq.~\eqref{mu_n}, which is
actually the energy of the linearized system.

To quantitatively analyze the impact of different $\gamma$ on the total energy 
of the vortex state, we utilized the linear solutions given by Eqs.~\eqref{m=-n}
and \eqref{R} as an approximation for the nonlinear system. 
This approach yielded the energy induced by inter-component interactions 
for $n \geq 1$ as expressed by the equation:
\begin{equation}
        E_\text{xpm}=2\pi\gamma\int_{0}^\infty |R^{n}_1(r)|^2|R^{n}_2(r)|^2rdr
        = \frac{\gamma C_{2n}^n}{2^{2n+3}\pi}.
    \label{Expm}
\end{equation}
Additionally, for the case where $n = 0$, the energy $E_\text{xpm}$ is equal to 0.
For the situations we discussed above, namely, $n =0, 1, 2$ and 3, the corresponding 
$E_\text{xpm}= 0$, $0.020\gamma$, $0.015\gamma$ and $0.012\gamma$, respectively.
This result indicates that the total energy of the system remains very close under 
different $\gamma$ values.

\begin{figure}[tbp]
\centering
\includegraphics[width=3.4in]{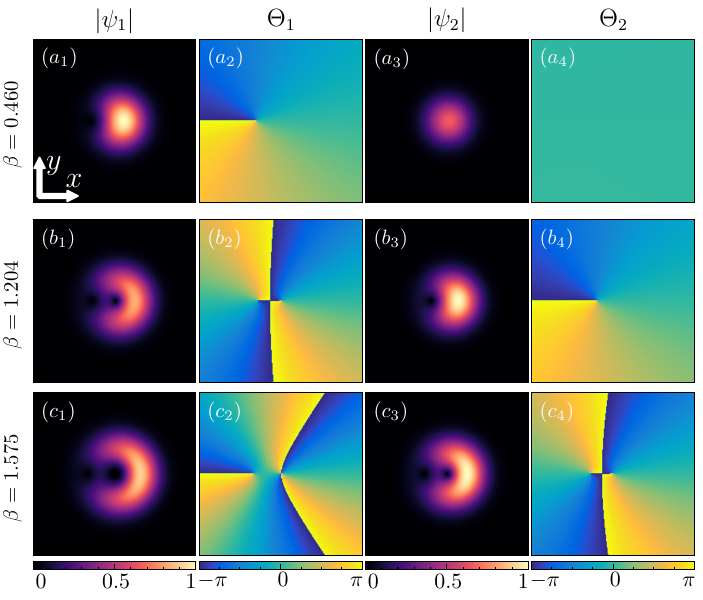}
\caption{Distributions of the absolute values, $\left\vert \protect\psi %
_{1,2}\right\vert $, and phases, $\Theta _{1,2}$, of wave functions of the
two components in the GS of the mixed-mode type for $\protect\beta =0.460$
(panels $a_{1}-a_{4}$), $\protect\beta =1.204$ (panels $b_{1}-b_{4}$), $%
\protect\beta =1.575$ (panels $c_{1}-c_{4}$), and $\Delta =0$, $\protect%
\gamma =-2$ in Eq. (\protect\ref{psi12}).}
\label{figure5}
\end{figure}

As concerns the (semi-) vortex GSs displayed in Fig. \ref{figure4} for $%
\gamma =1$, those for $\beta =0.3$, $1.0$, and $1.5$ correspond, severally,
to values of the angular momentum (\ref{momentum}) $M=0$, $-1$, and $-2$, as
is seen from comparison with Fig.~\ref{figure3}. These properties are
readily explained by the fact that in all these cases an absolute majority
of atoms belong to component $\psi _{1}$, which has the same values of
intrinsic vorticities, \textit{viz}., $m=0$,$-1$, and $-2$, respectively.
Thus, selecting appropriate values of $\beta $, it is possible to adjust the
vortex GS so as to realize any desirable value of its winding number.

On the other hand, the set of mixed-mode GSs found in the case of the
inter-component attraction ($\gamma =-2$), which are displayed in Fig.~\ref%
{figure5}, have half-integer values of angular momentum (\ref{momentum}),
\textit{viz}., $M(\beta =0.460)=-0.5$, $M(\beta =1.204)=-1.5$, and $M(\beta
=1.575)=-2.5$. These values of $M$ are explained by the fact that the
corresponding mixed states are composed of two components with vorticities $%
m\leq -1$ and $m+1$, which have equal weights (this is a generic property of
mixed-mode solitons in the free 2D space). Further, particular panels in
Fig.~\ref{figure5} demonstrate an essential difference of the mixed-mode GSs
from their counterparts of the (semi-)vortex type: in the mixed modes,
higher-order vortices with $|m|\geq 2$ tend to split in lower-order ones,
with separated pivots, and the pivots shift sidewise. The evolution of the
mixed-mode GSs can be summarized as follows: as $\beta $ increases, the next
phase transition increases the winding number of the right vortex in the $%
\psi _{1}$\ component by $1$, while $\psi _{2}$ assumes a shape similar to
that featured by the mixed-mode's $\psi _{1}$ component at the previous
stage.

\begin{figure}[tbp]
    \centering
    \includegraphics[width=3.4in]{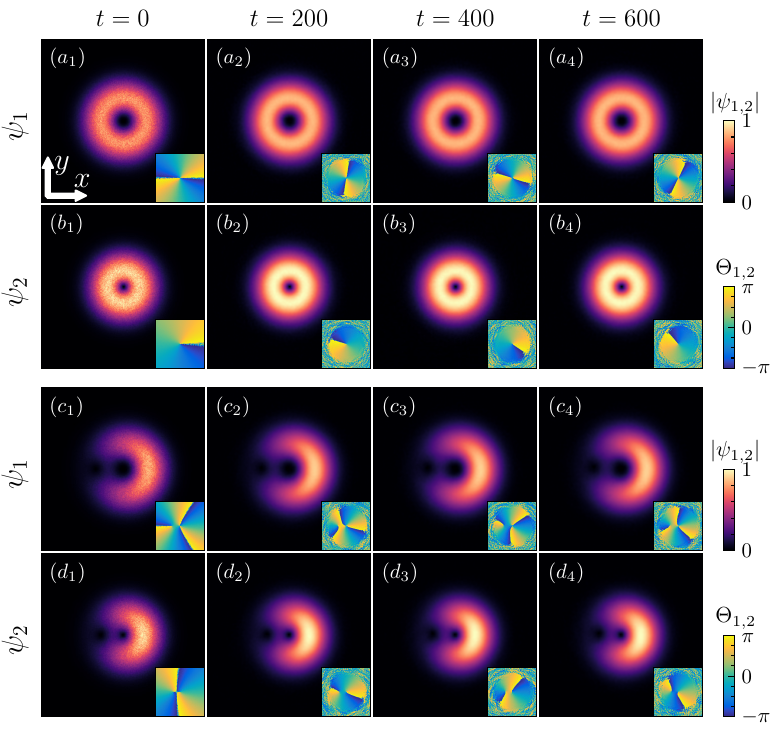}
    \caption{Numerical evolution of the perturbed ground state, where ($a_1-a_4$) 
    and ($b_1-b_4$) correspond to vortex state with parameters $\beta=1.5$,
    $\gamma=1$, while ($c_1-c_4$) and ($d_1-d_4$) correspond to mixed state 
    with parameters $\beta=1.575$, $\gamma=-2$. In each figure, the bottom right 
    corner depicts the phase distribution of the wave function. From left to right, 
    each column corresponds to the wave function density and phase distribution 
    at evolution times $t=0$ (initial state), $t=200$, $t=400$, and $t=600$.}
    \label{figure6}
\end{figure}

To corroborate that the GSs, identified by the above analysis, are indeed 
stable modes, as they should be, their stability was verified by real-time 
simulations of their perturbed evolution. We added white noise to the ground 
state wave functions, with the maximum noise intensity reaching $10\%$ of the 
wave function's amplitude. Subsequently, real-time evolution was applied 
to the perturbed wave functions. The results of numerical simulations are 
presented in Fig.~\ref{figure6}, showcasing a vortex state with parameters 
$\beta=1.5$, $\gamma=1$, and a mixed state with parameters $\beta=1.575$, 
$\gamma=-2$. It can be seen that at $t=0$, the wave functions exhibit numerous 
noise points. However, as time progresses, by $t=200$, the noise points on 
the wave functions disappear. Furthermore, in the subsequent evolution, the 
phases of the wave functions rotate around the vortex centers, while the 
amplitude distribution no longer undergoes changes. The results completely 
verify the stability of the ground states in all cases.

\section{The (semi-)vortex and mixed states as \textit{baby\ skyrmions}}

The realization of SOC in the two-component Bose gas\ suggests that it can
be considered as a (pseudo-)spin system, with the spin vector density
defined as
\begin{equation}
\mathbf{S}=\left( \psi ^{\dagger }\psi \right) ^{-1}\psi ^{\dagger }\mathbf{%
\sigma }\psi .  \label{vector}
\end{equation}%
For the vortex states with winding number $m\neq 0$ [recall $m=-n$, as in
Eq. (\ref{m=-n})], vector $\mathbf{S}$ produced by the linear solution~%
\eqref{R} is
\begin{equation}
\mathbf{S}(r,\theta )=\frac{1}{r^{2}+1}\left\{ 2r\cos \theta ,2r\sin \theta
,r^{2}-1\right\} .  \label{S}
\end{equation}%
For $m=0$, the spin-vector structure is trivial, $\mathbf{S}(r,\theta
)=\left\{ 0,0,1\right\} $.

\begin{figure}[tbp]
\centering
\includegraphics[width=3.4in]{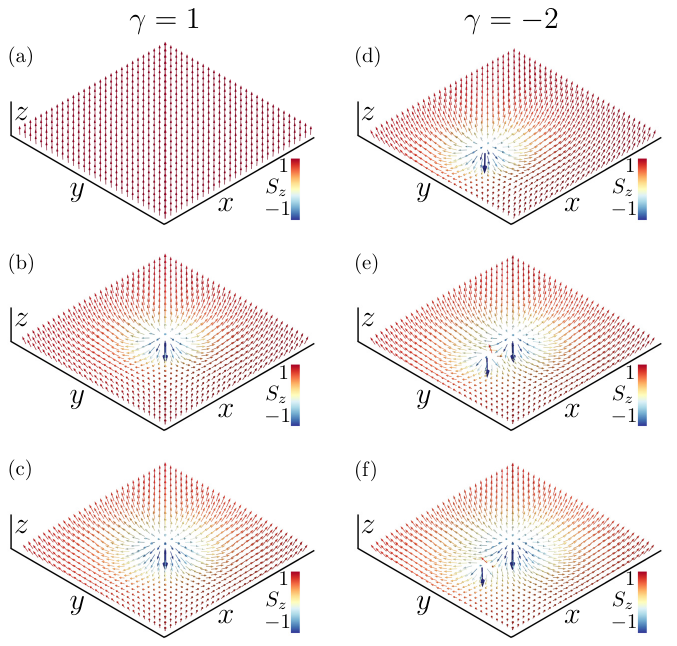}
\caption{Spin textures of the numerically constructed GSs of the (a-c)
vortex and (d-f) mixed-mode states in the nonlinear system with $\protect%
\gamma =1$ and $\protect\gamma =-2$, respectively. The vortex GSs in panels
(a-c) are identical to those displayed in Fig.~\protect\ref{figure4}, with
the same parameters, \textit{viz}., $\protect\beta =0.3$ and vorticities of
the two components $\left( 0,+1\right) $ (a), $\protect\beta =1.0$ and
vorticities $\left( -1,0\right) $ (b), and $\protect\beta =1.5$ and
vorticities $\left( -2,-1\right) $ (c). Similarly, the mixed-mode GSs in
panels (d-f) are identical to those displayed in Fig.~\protect\ref{figure5},
with the same parameters, \textit{viz}., $\protect\beta =0.460$ (a), $%
\protect\beta =1.204$ (b), and $\protect\beta =1.575$ (c). Arrows indicate
the direction of the spin vector (\protect\ref{vector}), and their colors
represent the magnitude of the $S_{z}$ component. Accordingly, the
respective directions vary from vertical-up (red) to vertical-down (blue)
ones. Bold blue arrows indicate the cores of the vortices.}
\label{figure7}
\end{figure}

The spin textures of the vortex states with winding numbers $m=0,-1$ and $-2$
in the $\psi _{1}$ component are displayed in Fig.~\ref{figure7}(a-c). These
textures for the states with $m\neq 0$ resemble those for 2D skyrmions,
which are often called \textquotedblleft baby skyrmions" (to stress the
difference from their full-fledged 3D counterparts). The \textquotedblleft
babies" are well-known nonlinear modes \cite{Bogolyub,Piette,Yasha,baby-opt}
which, in particular, were recently created in BEC \cite{baby-BEC}.

According to Eq.~\eqref{S}, $\lim_{r\rightarrow \infty }\mathbf{S}=\left\{
0,0,1\right\} $, hence the skyrmionic spin textures are embedded in the
asymptotically uniform vector background. Thus, the 2D space $\mathbb{R}^{2}$
may be compactified into the 2D sphere $S^{2}$. Note that spin vectors (\ref%
{S}) also take their values on $S^{2}$, therefore the skyrmion realizes the
second homotopy group $\pi _{2}(S^{2})=\mathbb{Z}$, characterized by the
topological number
\begin{equation}
Q=\frac{1}{4\pi }\int_{-\infty}^{\infty}\int_{-\infty}^{\infty}\mathbf{S}\cdot \left( \frac{\partial \mathbf{S%
}}{\partial x}\times \frac{\partial \mathbf{S}}{\partial y}\right) dxdy.
\label{Q}
\end{equation}%
The skyrmion topological number counts the number of times that $%
S^{2}$ is covered by the vector field $\mathbf{S}(r,\theta )$. The
substitution of expression~\eqref{S} in Eq.~\eqref{Q} yields $Q=-1$, which
is the topological charge of all the vortex states with $m\neq 0$, produced
by the linear and nonlinear versions of Eq.~(\ref{psi12}) alike. In
particular,\ the numerical calculation of topological charge $Q$ according
to Eq.~(\ref{Q}), for states displayed in panels (a-c) in Fig.~\ref{figure7}%
, yields $Q(m=0)=4.049\times 10^{-4},Q(m=-1)=-0.978$, and $Q(m=-2)=-0.962$,
respectively, in agreement with what is said above. Here the
numerical-integration domain is $0\leq r<8.0,0\leq \theta <2\pi $. According 
to Eq.~\eqref{Q}, numerical integration over the entire space, i.e., 
$-\infty<x,y<\infty$, is inherently challenging, and there is always a 
presence of systematic error. Increasing the integration range, while 
maintaining a constant spacing between the sampled points ($dxdy$), would 
indeed yield topological numbers $Q$ closer to integers. However, this 
approach necessitates significantly higher computational resources.

The mixed-mode states can be considered as a (nonlinear) superposition of
two baby-skyrmions with topological charges $Q=0$ and $-1$. The spin
textures of the mixed modes are presented in Figs.~\ref{figure7} (d-f), and
the respective numerically calculated values of topological charge (\ref{Q})
are $Q=-0.980,-1.931$ and $-1.891$, respectively. The corresponding values
for the ideal solutions are $Q=-1,-2$, and $-2$, respectively.

To visualize the structure of the skyrmions, it is relevant to identify
location of vortex core(s) in them, i.e., points where vector (\ref{S}) take
the value $\mathbf{S}=\left\{ 0,0,-1\right\} $. The vortex states with $%
m\neq 0$, displayed in Figs. \ref{figure4} and \ref{figure7}(a-c), feature
the single core, pinned to the center, $r=0$. On the other hand, Figs.~\ref%
{figure5} and \ref{figure7}(d-f) show that (as is actually mentioned above)
for the first mixed-mode state, the core position is offset from the center,
and two cores are featured by the second and third mixed modes.

\section{Conclusion}

In this work, we have reported results of the systematic analysis of vortex
states in the 2D SOC (spin-orbit-coupled) BEC under the action of the
gradient magnetic field and HO trapping potential. We have obtained exact
solutions for the linearized system and found that, by varying the SOC
strength and the magnetic-field gradient, the energy-level inversion can be
realized, allowing the trapped higher-order vortex modes to transition into
the GS (ground state). We have also solved the full nonlinear system
numerically and found that, in the case of the inter-component repulsion,
the results for the vortex modes are consistent with the predictions of the
linear theory. On the other hand, the attractive inter-component interaction
creates mixed-mode states in the vicinity of the GS phase-transition points.
Then, we have analyzed the GS spin textures and found that both the vortex
and mixed-mode states have the structure of the 2D (\textit{baby})
skyrmions. Our findings predict possibilities for the creation of stable
higher-order vortex states in experiments with BEC.

It may be relevant to elaborate similar settings, emulating the SOC and the
action of gradient magnetic field, in optics. A challenging possibility is
to extend the analysis for 3D systems, cf. Ref. \cite{Han Pu}.

\section*{Acknowledgments}

This work was supported by Natural Science Foundation of Guangdong province
through grant No. 2021A1515010214, NNSFC (China) through grants Nos.
12274077, 11905032, and 11904051, the Research
Fund of the Guangdong-Hong Kong-Macao Joint Laboratory for Intelligent
Micro-Nano Optoelectronic Technology through grant No. 2020B1212030010, and
Israel Science Foundation through Grant No. 1695/22.

\end{document}